\documentclass[aps,pra,reprint,superscriptaddress,longbibliography]{revtex4-1}

\usepackage{amsmath}
\usepackage{xcolor}
\usepackage{textcomp}
\usepackage{graphicx} 
\usepackage{float}
\usepackage{epstopdf}
\usepackage[colorlinks,citecolor=blue,linkcolor=blue]{hyperref}

\newcommand{\CNN}{Centre de Nanosciences et de Nanotechnologies, CNRS, Université Paris-Saclay, Palaiseau, France}
\newcommand{\DEE}{Department of Electrical Engineering and ICT, University of Naples Federico II, Naples, Italy} 
 
\newcommand{\ITEFI}{Instituto de Tecnologías Físicas y de la Información (CSIC), Madrid, Spain} 
\newcommand{\LabSTICC}{LabSTICC, CNRS, Universit\'e de Bretagne Occidentale, Brest, France}
\newcommand{\LAF}{Laboratoire Albert Fert, CNRS, Thales, Université Paris-Saclay, Palaiseau, France}
\newcommand{\SPEC}{SPEC, CEA, CNRS, Université Paris-Saclay, Gif-sur-Yvette, France}

\begin{document}

\title{Nonlinear mode interactions under parametric excitation in a magnetic microdisk}

\author{G. Soares}
\affiliation{\SPEC}
\author{R. Lopes Seeger}
\affiliation{\SPEC}
\affiliation{\CNN}
\author{A. Kolli}
\affiliation{\SPEC}
\author{M. Massouras}
\affiliation{\CNN}
\author{T. Srivastava}
\affiliation{\CNN}
\author{J.-V. Kim}
\affiliation{\CNN}
\author{N. Beaulieu}
\affiliation{\LabSTICC}
\author{J. Ben Youssef}
\affiliation{\LabSTICC}
\author{M. Mu\~{n}oz}
\affiliation{\ITEFI}
\author{P. Che}
\affiliation{\LAF}
\author{A. Anane}
\affiliation{\LAF}
\author{S. Perna}
\affiliation{\DEE}
\author{M. d'Aquino}
\affiliation{\DEE}
\author{C. Serpico}
\affiliation{\DEE}
\author{H. Merbouche}
\email{hugo.merbouche@cea.fr}
\affiliation{\SPEC}
\author{G. de Loubens}
\email{gregoire.deloubens@cea.fr}
\affiliation{\SPEC}

\begin{abstract}

  A pair of quantized spin-wave modes is driven by two-tone parallel pumping in a magnetic microdisk, thus realizing an unexplored yet simple spectroscopic method for studying their nonlinear interaction. The nonlinear dynamics is experimentally investigated by probing the resulting steady state, which is found to critically depend on the chosen pair of modes, the detuning between the pump frequencies and the modes parametric resonance, as well as on the temporal sequence of the two rf tones. A general theory of parametric excitation in confined structures based on magnetization normal modes quantitatively accounts for the observed dependence and non-commutative behaviors, which emerge from the interplay between the self and mutual nonlinear frequency shifts of the spin-wave modes. Owing to its high degree of external controllability and scalability to larger sets of modes, this dynamical system provides a model platform for exploring nonlinear phenomena and a promising route toward rf driven state mapping relevant to unconventional computing.

\end{abstract}

\maketitle

\section{Introduction}

As a highly nonlinear system, spin-waves (SWs) are a promising platform for uconventional computing \cite{hughes19,markovic20,papp21,koerber23}. In extended magnetic films, the magnon spectrum is continuous and often degenerate, leading to complex phenomena like BEC \cite{demokritov06,schneider20,divinskiy21a}, solitons \cite{kosevich90,slavin05c,mohseni13} instabilities and chaos \cite{gibson84,rezende90,petit-watelot12}. These phenomena often involve an intractable number of modes which makes them difficult to model and control. In contrast, in confined nanostructures, the SW spectrum is quantized and the number of modes involved in the nonlinear dynamics can be tracked \cite{li19c,schultheiss19,mohseni21,hamadeh23a,merbouche24a,ngouagniayemeli25a}. For instance, understanding the nonlinear dynamics of spin-torque nano-oscillators (STNOs), that can be described by a single-mode theory \cite{slavin09}, greatly helped optimizing their properties for rf applications \cite{houssameddine07,urazhdin10a}, sensing \cite{fang16,wittrock24} and neuromorphic computing \cite{torrejon17,romera18,markovic19,zahedinejad20}, where arrays of coupled STNOs are used to classify rf inputs. Recently, it was demonstrated that classification tasks could also be achieved using the different SW modes of a single magnetic microdisk \cite{koerber23}. In this approach, the rf inputs are inherently interconnected in the reciprocal space through the nonlinear SW modes interactions \cite{verba21}, alleviating the need for physical interlinks and opening the possibility to program reconfigurable neural-like computing architectures using rf signals.

\begin{figure}
	\centering
        \includegraphics[width=\columnwidth]{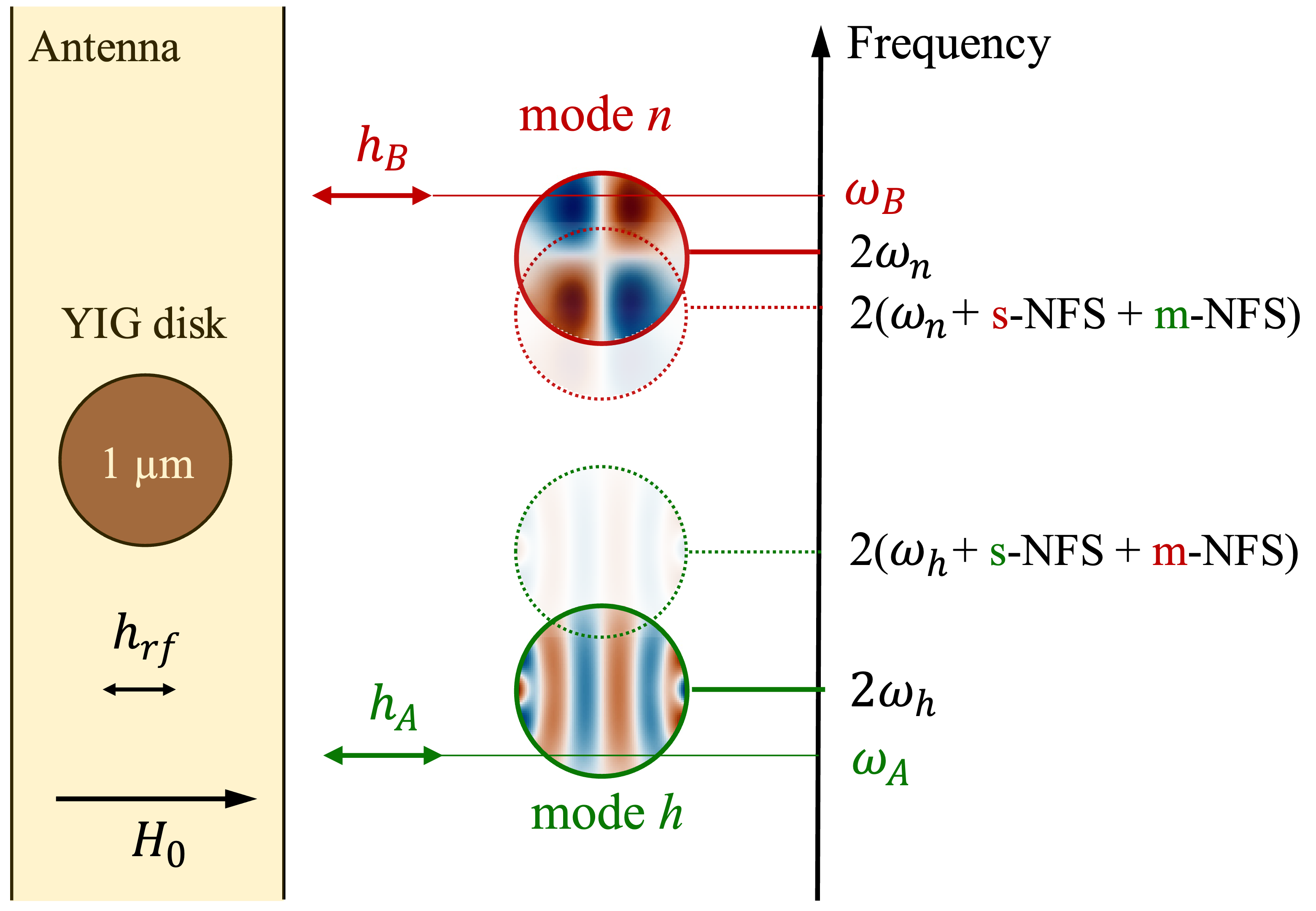}
	\caption{Schematics of the experiment. An rf antenna is patterned on top of a 52 nm thick YIG disk magnetized in-plane and generates an rf pumping field parallel to the static field. Injecting two rf signals with frequencies $\omega_A$ and  $\omega_B$, it is possible to selectively excite two distinct SW modes simultaneously. The two modes nonlinearly interact via their self and mutual nonlinear frequency shifts (s-NFS, m-NFS) and the steady state is reached when both are detuned from their respective pump by a certain critical amount.}
	\label{fig:1}
\end{figure}

In this study, we aim at determining the nonlinear phenomena involved when multiple modes are excited in a confined magnetic system, as well as the ways to model and control them. To that end, we use parallel pumping \cite{gurevich96a,ulrichs11a,guo14,braecher17b} which allows us to selectively excite any SW mode, as we have shown previously \cite{srivastava23a}. In the first part, we apply a single rf frequency which excites a single SW mode in a disk, whose intensity in the steady state is recorded. The results of this single-tone spectroscopy are found in good agreement with a general theory \cite{aquino26} describing parallel pumping in nanostructures using the normal modes model (NMM) approach \cite{aquino09,perna22}, which expands the nonlinear Landau-Lifshitz-Gilbert (LLG) equation on the eigenbasis of the SW modes of the disk, thereby going beyond the macrospin ansatz used in Ref.~\cite{guo14}. In the second part, we apply two pumping frequencies ($\omega_A$ and  $\omega_B$), each exciting a single mode. In this way, any pair of SW modes can be pumped simultaneously in the microdisk (see Fig.~\ref{fig:1}). This two-tone spectroscopy reveals a non-trivial steady-state dynamics, which crucially depends on the pair of modes and on excitation conditions. In particular, we evidence non-commutative behaviors, where the final state depends on the temporal sequence of rf inputs ($\omega_A$ applied before $\omega_B$ or vice versa). Using the theory based on normal modes, we unveil that the dominant nonlinear terms controlling the dynamics are the self and mutual nonlinear frequency shifts (s- and m-NFS) of the SW modes, \textit{i.e.}, the fact that each mode’s frequency depends on its own as well as on the other mode’s amplitude. The developed theory details the computation of eigenmodes and relevant resonant nonlinear terms, derives the conditions for the onset of parametric instability, and exhaustively treats all the cases where two modes are simultaneously driven into instability by a two-tone excitation to describe the resulting steady states \cite{aquino26}. In the present paper, we use it to reproduce our specific experimental results and to explain how the observed steady states depend on the signs of the NFS of the two involved modes. Thanks to the high sensitivity of the dynamics to the detuning between the pump frequencies and the modes parametric resonance, our experimental method allows for a spectroscopic determination of these important nonlinear parameters.

\section{Results}

\subsection{Sample and experimental set-up}

A 52~nm thick yittrium iron garnet (YIG) film grown on GGG (111) by liquid phase epitaxy is patterned into a 1~µm diameter disk. The Gilbert damping of the disk is measured to be $5 \times 10^{-4}$. A 3~µm wide gold antenna is patterned on top of the disk and generates a uniform excitation field that is parallel to the static magnetic field $H_0$ (in the following, it is fixed to $\mu_0H_0=27.1$~mT). This parallel pumping field couples to the longitudinal component of the magnetization due to the elliptical character of the precession. Due to the nanostructuring, the SW modes are quantized with their resonance frequencies spaced by few tens of MHz. When the frequency of the pumping field is about twice the frequency of a SW mode, and its amplitude is above a critical threshold, the pumping compensates the mode's relaxation and the mode's amplitude grows exponentially: it undergoes parametric instability. Using a multi-channel pulsed microwave generator and a combiner, multiple rf frequency tones can be applied simultaneously to the antenna so that each can excite a selected SW mode (an rf power of 0~dBm corresponds to an rf field of $0.3\pm0.1$~mT). The magnetization dynamics is detected using a magnetic resonance force microscope (MRFM, see Appendix~\ref{sec:MRFM} for details) \cite{guo14,srivastava23a} which yields a signal proportional to the SW intensity induced by the microwave signals \cite{klein08}.

\subsection{Single-tone experiment and modeling}

\begin{figure}
	\centering
        \includegraphics[width=\columnwidth]{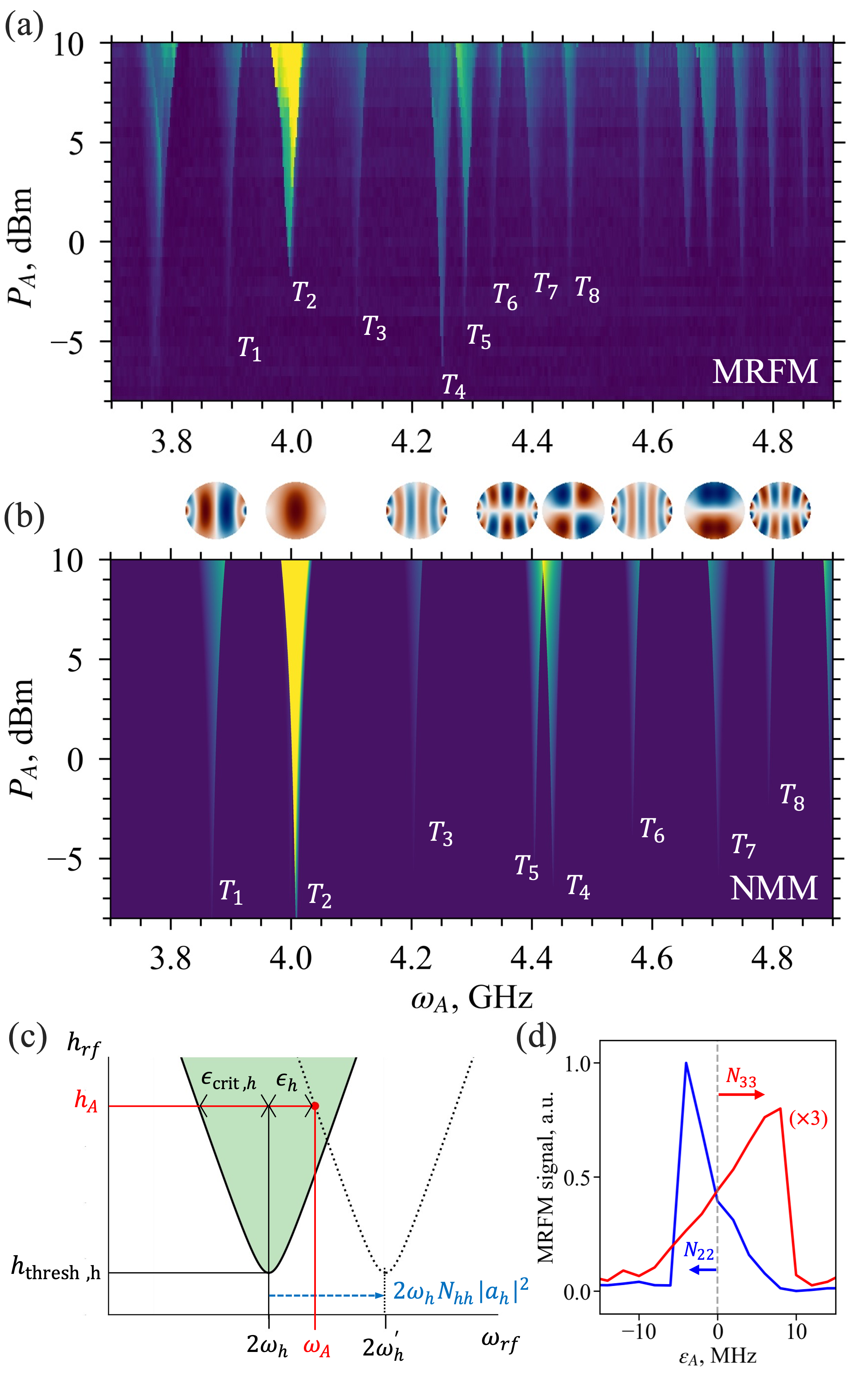}

	\caption{Parametric spectroscopy of a 1~µm YIG disk: (a) experimental MRFM data, (b) simulation data using the NMM. Individual SW modes (top of b) are selectively excited at twice their resonance frequency by the parallel pumping field. (c) Sketch of the excitation condition for a single mode. The steady state is reached when the mode is critically detuned from the pump frequency, \textit{i.e.}, when its parametric resonance frequency has shifted by the available detuning $(\epsilon_{A,h}+\epsilon_{\text{crit},h})$. (d) Experimental normalized steady-state intensities of $T_2$ and $T_3$ with opposite slopes due to the sign of their s-NFS.}
	\label{fig:2}
\end{figure}

We start by applying a single rf tone to the antenna. The microwave is pulse-width-modulated (PWM) at the frequency of the MRFM cantilever: pulses are 38~µs long with a 50\% duty cycle, guaranteeing that the magnetization has time to fully relax to thermal equilibrium between each pulse. Given the characteristic times, the MRFM amplitude is mostly sensitive to the SW intensity excited by the microwave in the steady state. The MRFM signal versus the pumping field frequency and power is shown in Fig.~\ref{fig:2}(a). As reported previously \cite{srivastava23a}, the parallel pumping allows to selectively excite the quantized SW modes of the disk. The MRFM data displays a collection of typical tongue shape instability regions \cite{ulrichs11a,guo14}. The tongues are numbered from $T_1$ to $T_8$. The steady-state intensities of the excited modes are quite uniform even for higher order modes (note the linear color scale). For certain modes, the MRFM amplitude is maximal on the right edge at the bottom of the tongues ($T_1$, $T_3$, $T_5$, $T_6$, $T_8$), while for some others it is the case on the left edge ($T_2$, $T_4$, $T_7$).

These features are reproduced using the analytical model based on the NMM as shown in Fig.~\ref{fig:2}(b), allowing us to associate the experimental tongues to the analytical ones. The micromagnetically computed spatial profiles (see Appendix~\ref{sec:micromag}) of the SW modes associated with each tongue are displayed on top of the graph. Note that the experimental tongue below $T_1$ corresponds to edge modes which are notoriously difficult to model \cite{guo13,duan15}, therefore we will not comment on it further in the following.

Let us consider a single mode with eigenfrequency $\omega_h$ pumped by a longitudinal rf field of frequency $\omega_A$ and amplitude $h_A$. Following the derivation in \cite{aquino26}, the well-known parallel pumping equation is recovered \cite{gurevich96a,braecher17b}, with $a_h$ the amplitude of mode $h$:
\begin{equation} \label{eq:1t-pp}
    \dot{a_h}=(j\omega_h-\lambda_h)a_h+2jV_{h}h_A\cos{(\omega_At)} a_h^* \,,
\end{equation}
where $\lambda_h$ is the relaxation frequency (inverse of the lifetime) and $V_h$ (in units of the gyromagnetic ratio $\gamma$) is the coupling coefficient of mode $h$ to the longitudinal rf field ($=0$ if the precession is circular). These two parameters can be computed directly from the spatial profile of the mode \cite{aquino26}. This equation describes that mode $h$ will undergo parametric instability if the red point $(\omega_a,h_a)$ is within the green area in the $(\omega_{rf},h_{rf})$ plane (Fig.~\ref{fig:2}(c)), corresponding to an Arnold tongue shape in the $(\omega_A,P_A)$ plane in Figs.~\ref{fig:2}(a,b). The Arnold tongue is described by two conditions: (i) the pumping field magnitude must be larger than a certain threshold field $h_{\text{thres},h}$, and (ii) the absolute value of the detuning $\epsilon_{A,h}$ between the pump frequency and twice the mode frequency $\epsilon_{A,h}=\omega_A-2\omega_h$ is smaller than a certain critical detuning  $\epsilon_{\text{crit},h}$ which depends on the magnitude of the pumping field:
\begin{subequations}
\begin{eqnarray}
    h_A > h_{\text{thres},h} &~;~& |\epsilon_{A,h}| = |\omega_A - 2 \omega_h| < \epsilon_{\text{crit},h}
    \,,\\
    h_{\text{thres},h} = \frac{\lambda_h}{|V_h|} &~;~& \epsilon_{\text{crit},h} = 2|V_h|\sqrt{h_A^2 - h_{\text{thres},h}^2}
    \,.
\end{eqnarray}
\end{subequations}
Outside the Arnold tongue, the pump is too detuned from the parametric resonance $2\omega_h$ and there is no excitation. Inside the instability region, the mode amplitude grows exponentially at a rate $\Gamma>0$. At the edge, on the critical (black) line $\epsilon_{A,h}=\pm\epsilon_{\text{crit},h}$, $\Gamma=0$.

In our confined geometry, the mechanism responsible for the saturation of the mode intensity in the tongue is the s-NFS $N_{hh}$, describing the shift of the mode frequency as its intensity grows, 
\begin{equation}
    \omega_h' = \omega_h(1+N_{hh}|a_h|^2) \,. 
\end{equation}
It is a unitless quantity originating from the resonant 4-magnon scattering terms of the mode interacting with itself \cite{aquino26,lvov94}. The mode intensity can grow as long as its frequency $\omega_h'$ is not too detuned from the pump (red point of coordinates $\omega_A,h_A$ in Fig.~\ref{fig:2}(c)). The steady state is reached when the point $(\omega_A,h_A)$ is at the edge of the shifted tongue for which $\Gamma=0$ (dotted black line). The steady-state intensity $|a_h|^2_\pm$ thus writes
\begin{equation} \label{eq:stab}
  2\omega_hN_{hh}|a_h|^2_\pm = \epsilon_{A,h} \pm \epsilon_{\text{crit},h} \,, \text{with }\pm = \text{sign}(N_{hh}) \,.
\end{equation}
It is reached when the mode parametric resonance has shifted by its maximum available detuning, which linearly varies with $\omega_A$ between $0$ and $\epsilon_{\text{crit},h}$ between the edges of the tongue. This results in the typical saw-tooth shape \cite{guo14} of the steady-state intensity of the mode found in the experiments (see Fig.~\ref{fig:2}(d)) for modes with positive and negative s-NFS. The smaller the NFS, the higher the mode can grow before it is critically detuned, which explains the observed differences of steady-state amplitudes in our experiment.

One can observe that certain modes shift up ($T_1$, $T_3$, $T_5$, $T_6$, $T_8$) while others shift down ($T_2$, $T_4$, $T_7$) in Fig.~\ref{fig:2}(a). This is a direct consequence of the confined geometry, as the NFS is purely negative for a full film magnetized in-plane. This behavior is well taken into account by our theory (see Fig.~\ref{fig:2}(b)), and the general argument boils down to the spatial localization of the mode in the inhomogeneous internal field of the disk and to the competition between the static (positive NFS) and dynamic contributions (negative NFS). The confined geometry also greatly influences the saturation mechanism. In the 1990’s, L’vov and co-workers developed the S-theory \cite{lvov94} and showed that in most experiments performed on extended films the saturation is caused by a phase mechanism which could be accounted for using the resonant 4-magnon scattering terms $T$ and $S$ from the continuous ensemble of SW modes resonant with the pump. In our confined geometry, the mode spacing (tens of MHz) is much larger than the critical detuning (few MHz), which guarantees that only a single SW mode is excited by the pump. In that case, the s-NFS becomes the critical mechanism that governs the steady-state intensity.

\subsection{Two-tone experiment}

We now investigate the interaction between pairs of parametrically excited modes by introducing a second rf tone in the antenna. The two microwave signals are PWM with a 3~µs delay to ensure that the mode excited first has time to reach its steady-state amplitude before the application of the second tone.

\begin{figure}
	\centering
        \includegraphics[width=\columnwidth]{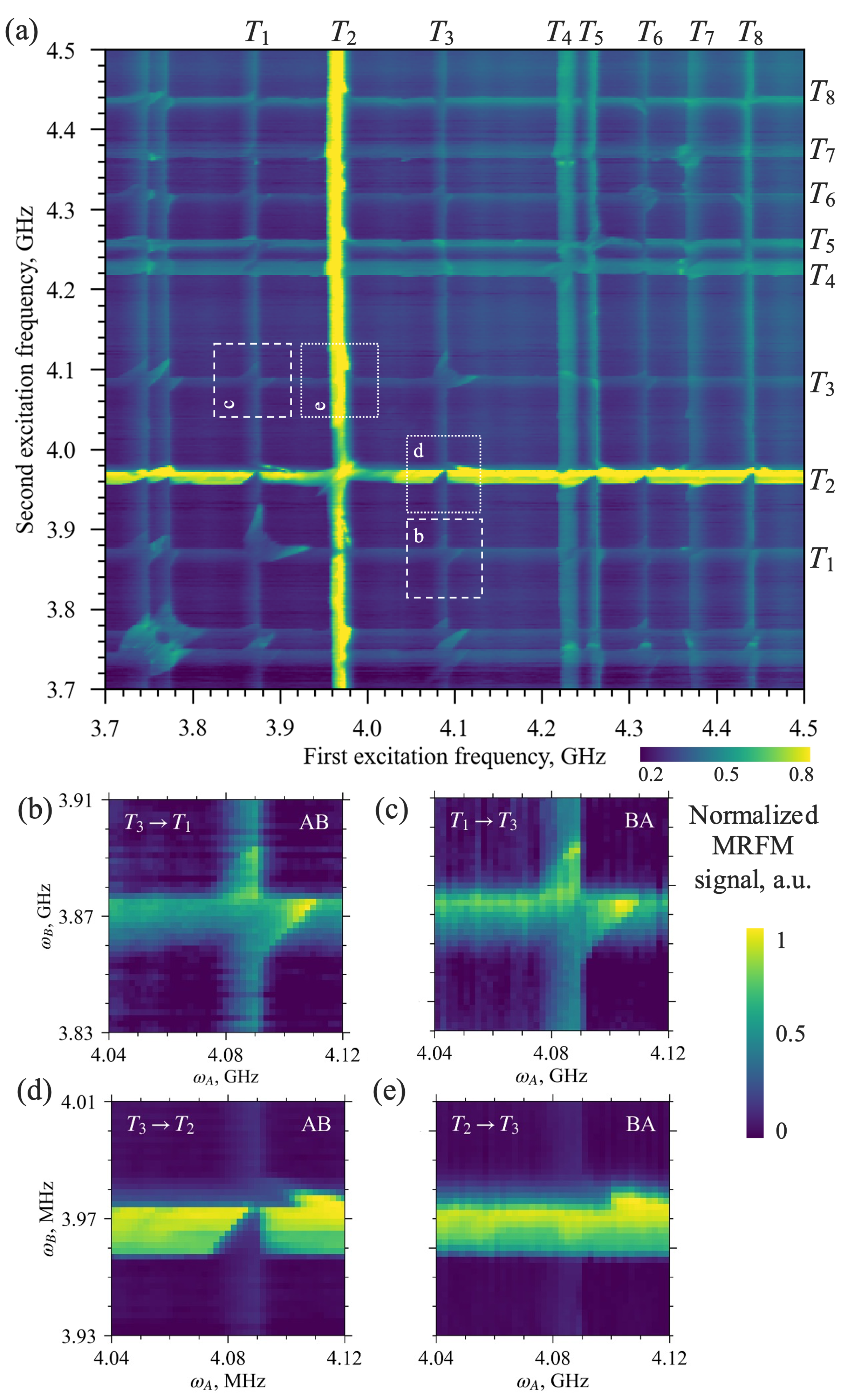}

	\caption{(a) Two-tone parametric spectroscopy showing the excitation of pairs of SW modes ($T_1$ to $T_8$) and recording the total SW intensity in the steady state as a function of the two pumping frequencies ($P_A = P_B = 4$~dBm). (b,c,d,e) Zoom on four typical crossings. Pumping frequency $\omega_A$ addresses $T_3$, while $\omega_B$ addresses $T_1$ (b,c) or $T_2$ (d,e).  In the left (right) column, $\omega_A$ ($\omega_B$) is introduced first: AB (BA). $T_3$-$T_1$ exhibits commutativity while $T_3$-$T_2$ is non-commutative.}
	\label{fig:3}
\end{figure}

The two-tone spectroscopy is presented in Fig.~\ref{fig:3}(a). The bright horizontal and vertical stripes correspond to frequencies for which a mode is excited. Each stripe is associated and labelled with the corresponding tongue from the single-tone spectroscopy (Fig.~\ref{fig:2}(a)). The vertical stripes correspond to the tone that is applied first. In the crossing areas, where two stripes intersect, two distinct modes are excited simultaneously (except on the diagonal, where both frequencies address the same mode). The steady-state intensity at the crossings does not correspond to the sum of the respective single tone intensities, meaning that the modes interact nonlinearly. Moreover, a wide variety of intensity patterns are observed at the different crossings. Certain patterns seem to be shared by multiple pairs of mode, suggesting a common underlying mechanism. The intensity at the mode crossings also strongly depends on which tongue is excited first. For instance, when $T_2$ is excited first (bright vertical stripe at 3.97~GHz), it is mostly undisturbed when crossing other tongues. However, when it is applied second (bright horizontal stripe at 3.97~GHz), it is strongly suppressed in small triangular areas when it crosses $T_1$, $T_3$, $T_5$, $T_6$ and $T_8$, which are the modes that have positive NFS. For these modes, the final state depends on the temporal sequence of excitation: the interaction is labelled as non-commutative. On the contrary, when $T_3$ (at 4.08~GHz) crosses $T_1$ and $T_6$, the intensity does not depend on the sequence: the interaction is commutative. This seems to indicate that the final state critically depends on the sign of the s-NFS, which is opposite for $T_2$ and $T_3$.

A zoom in on two typical crossings is shown in Figs.~\ref{fig:3}(b-e): one commutative ($T_3$-$T_1$) and one non-commutative ($T_3$-$T_2$). To better visualize the commutation, the pumping frequency $\omega_A$ associated with the tongue $T_3$ is always placed on the x-axis, while the frequency  $\omega_B$ associated with $T_1$ or $T_2$ is on the y-axis. For the graphs on the left column, $\omega_A$ is applied first (labelled AB), while on the right column $\omega_B$ is applied first (BA). The temporal sequence AB and BA are sketched in Fig.~\ref{fig:4}(a). Strikingly, $T_3$ interacts completely differently with $T_1$ and $T_2$. The final state also strongly depends on the pumping frequencies  $\omega_A$, $\omega_B$ inside the tongues.

\subsection{Two-tone modeling}

\begin{figure*}
	\centering
        \includegraphics[width=\textwidth]{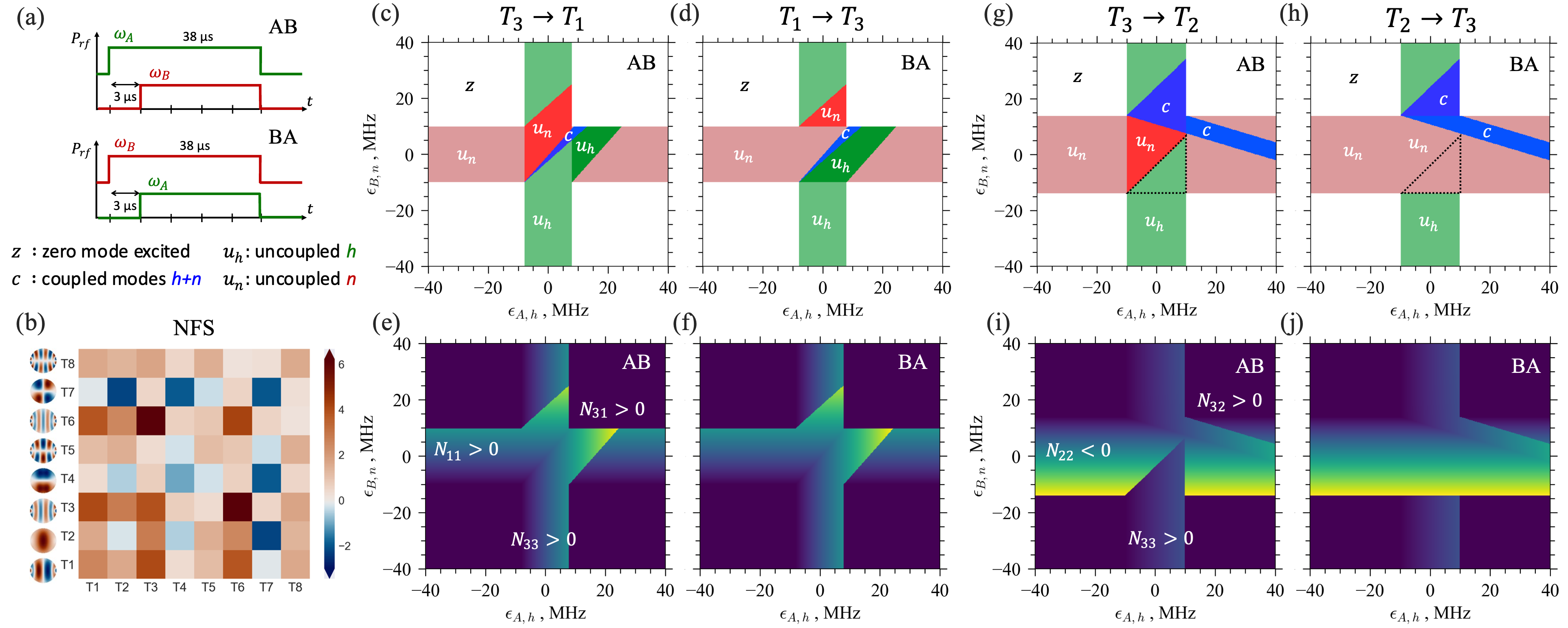}
	\caption{(a) Excitation scheme: the two pumps are PWM with frequency $\omega_A$ (resp. $\omega_B$) applied 3~µs after the other, labelled as AB (resp. BA). (b) Calculated s- and m-NFS for the eight modes observed in the experiment. (c,d) Phase diagrams representing the steady-state solution and (e,f) total steady-state intensity for the $T_3$-$T_1$ crossing, where all NFS are positive. (g,h) and (i,j) same for the $T_3$-$T_2$ crossing, where $N_{22}<0$ and  $N_{32}>0$. The NFS ratios are fitted from the experiment (vs. calculated ratios): $N_{33}/N_{31}=1.0\pm0.1~(0.97)$,  $N_{11}/N_{31}=1.27\pm0.1~(0.66)$, $N_{33}/N_{32}=0.95\pm0.1~(1.4)$ and $N_{22}/N_{32}=-0.5\pm0.1~(-0.1)$.}
	\label{fig:4}
\end{figure*}

Only one additional ingredient is necessary to model this wide diversity of final states: the m-NFS. Indeed, in the presence of two modes, mode $h$ (excited by $\omega_A$) and mode $n$ (excited by $\omega_B$), there are additional resonant 4-magnon scattering terms describing the linear variation of the frequency of mode $h$ with the mode intensity $|a_n|^2$ and vice versa:
\begin{subequations}
\begin{eqnarray}
  \omega_h' &=& \omega_h(1+N_{hh}|a_h|^2+N_{hn}|a_n|^2)
  \,,\\
  \omega_n' &=& \omega_n(1+N_{nn}|a_n|^2+N_{nh}|a_h|^2)
  \,.
\end{eqnarray}
\end{subequations}
It is interesting to note that the unitless coefficients $N_{hn}$ and  $N_{nh}$ are equal \cite{aquino26}. Their values computed for the modes in the tongues $T_1$-$T_8$ are displayed in Fig.~\ref{fig:4}(b). The matrix is symmetric, and the diagonal is composed of the s-NFS terms. Outside the diagonal, the composition rules are not trivial as they involve how one mode is sensitive to the local changes in the internal and dynamic field induced by each other modes. 

The set of coupled parametric pumping equations for any pair of modes is given by:
\begin{subequations} \label{eq:2t-pp}
\begin{eqnarray}
  \dot{a_h} &=& (j\omega_h'-\lambda_h)a_h+2jV_{h}h_A\cos{(\omega_At)} a_h^*
  \,,\\
  \dot{a_n} &=& (j\omega_n'-\lambda_n)a_n+2jV_{n}h_B\cos{(\omega_Bt)} a_n^*
  \,.
\end{eqnarray}
\end{subequations}
The existence and stability of the solutions to these equations have been rigorously derived in \cite{aquino26}. The method allows to build phase diagrams such as the ones shown in Fig.~\ref{fig:4} which display the steady-state solutions for every pair of excitation frequencies $(\epsilon_{A,h},\epsilon_{B,n})$ at the crossing $T_3$-$T_1$ (Figs.~\ref{fig:4}(c,d)) and the crossing $T_2$-$T_1$ (Figs.~\ref{fig:4}(g,h)). We can distinguish three main regions.
\begin{itemize}
\item If both pumping frequencies are too detuned from the two considered modes, then zero mode is excited and the steady state is trivial. These are the white areas outside the cross, labeled ``z'', for which $|\epsilon_{A,h}| > \epsilon_{\text{crit},h}(h_A)$ and $|\epsilon_{B,n}| > \epsilon_{\text{crit},n}(h_B)$. 
\item If only one pumping frequency satisfies the excitation conditions, then the only steady-state solution is composed of a single uncoupled mode, here labeled ``uncoupled mode $h$'': $u_h$ (light green area) or ``uncoupled mode $n$'': $u_n$ (light red area). The solutions $u_h$ and $u_n$ correspond to the single mode stability condition (\ref{eq:stab}) previously established.
\item If both modes are excited, then the steady state can be composed of one mode or two modes. The two coupled modes steady state is labeled as ``c'' (blue areas). The single mode solution $u_h$ (bright green area) or $u_n$ (bright red area) can occur when the second excited mode suppresses the first mode in the process described below. 
\end{itemize}

When the first exciting frequency is introduced, the first mode grows to its steady-state intensity. This linearly shifts the resonance frequency of the second mode by 0 at one edge of the first mode tongue and maximally at the opposite edge. This results in a tilting of the tongue of the second mode in the central part (bright colors areas in the phase diagram). In these characteristic tilted areas, two modes are excited. When the second frequency is introduced, the first mode is initially at steady state: it is critically detuned from its pump. Meanwhile the second mode will grow exponentially until it becomes critically detuned from its pump. The first mode can either benefit or suffer from the growth of the second mode. If the m-NFS is of the same sign as the s-NFS of the first mode, then the first mode gets detuned by the second mode, and its intensity will decrease, possibly until its suppression, for as long as the second mode has not reached steady state. If the NFS signs are opposite, then the first mode gets better tuned to its pump and its intensity increases. Ref.~\cite{aquino26} reports a general and quantitative treatment for all possible combinations of NFS coefficients.

For the $T_3$-$T_1$ crossing where $N_{11}$, $N_{33}$ and $N_{13}$ are all positive (Figs.~\ref{fig:4}(c,d)), both tongues are shifted toward the upper right corner. The steady-state solution is the same when the rf signals are commutated. Indeed, in most bright areas (both modes excited), the first mode is suppressed by the second, matching the light color areas (single mode excited) solution where they overlap. For the $T_3$-$T_2$ crossing (Figs.~\ref{fig:4}(g,h)), $N_{33}$, $N_{23}$ are positive, but now $N_{22}$ is negative. The vertical tongue associated with mode 3 is shifted to the bottom right corner (bright blue). As a result, the dashed triangle is single mode $h$ (light green) for AB and single mode $n$ (light red) for BA which explains the observed strong non-commutative behavior. 

From the stable solution phase diagram, we compute the total steady-state intensity (Figs.~\ref{fig:4}(e,f,i,j)). The ratio between the NFS coefficients have been fitted from the experiment. Although our model predicts their correct sign and order of magnitude, it is not a perfect match, which is to be expected since the experimental eigenfrequencies are not precisely aligned with the model. NFS ratios are particularly sensitive to the actual mode profile for modes with a s-NFS close to zero or with precession near the disk edges. Our two-tone spectroscopy method allows to precisely extract these nonlinear coefficients, which are found to vary substantially with the chosen pair of modes, and are the only parameters necessary to reproduce the wide variety of steady states, as evidenced by the remarkable agreement between the experimental and calculated maps of Figs.~\ref{fig:3} and \ref{fig:4}, respectively (see also other crossings presented in Appendix~\ref{sec:otherXings}).

\section{Discussion}

Our experiments demonstrate that the steady-state phase diagram is governed by the interplay between s- and m-NFS, pump detuning, and excitation timing. The complexity of two-mode interactions originates from the four-dimensional phase space of the dynamics, which gives rise to multiple coexisting stable steady states \cite{aquino26}. The delayed PWM excitations employed in the current experiments enable to reach some of them. Additional tunability via the pumping power and excitation protocols enables deterministic selection among the coexisting solutions. For instance, continuous-wave (CW) frequency sweeps allow nontrivial states to persist beyond the crossing of Arnold tongues. The emergence of quasiperiodic regimes associated with a loss of synchronization between SW modes and the driving tones is also predicted and numerically confirmed. These features are analyzed in detail within the general theoretical framework developed in Ref.~\cite{aquino26}. The nonlinear transient regimes, which we did not specifically address in this study, are also expected to exhibit interesting behaviors.

In conclusion, we experimentally investigated nonlinear interactions between arbitrary pairs of SW modes in a YIG microdisk under parallel pumping. Despite the apparent simplicity of the system, these interactions generate a rich nonlinear phase space characterized by multi-stability and strong sensitivity to excitation conditions. By projecting the dynamics onto the eigenmode basis, the observed behavior can be embedded within a reduced theoretical framework that is applicable to arbitrary geometries and magnetic ground states. Extension to larger sets of modes thanks to frequency multiplexing is expected to open up the possibility of chaotic dynamics, while the number of possible stable solutions should scale as $2^p$ where $p$ is the number of modes. Our findings therefore establish parametrically driven magnonic systems as a controllable platform for nonlinear dynamical phenomena. Furthermore, they enable the mapping of temporal rf inputs onto programmable steady states for classification and learning. Finally, we would like to emphasize that our approach could be applied to other dynamical systems where nonlinear interactions between different excitation modes play an important role, as in hydrodynamics \cite{kambe90} and opto-mechanics \cite{asadi18}.

\section*{Acknowledgements}

This work is supported by the Horizon2020 Research Framework Program of the European Commission under Grant No. 899646 (k-NET) and by the Agence Nationale de la Recherche under Grant No. ANR- 20-CE24-0012 (Marin).

\appendix

\section{MRFM set-up} \label{sec:MRFM}

The SW parametric spectroscopy of the YIG disk is performed using a magnetic resonance force microscope (MRFM), which is located between the poles of an electromagnet and operated under vacuum ($10^{-6}$~mbar) at room temperature \cite{klein08}. It uses a very soft cantilever (spring constant $k \simeq 6$~mN/m), at the end of which a cobalt spherical probe of diameter 500~nm is attached \cite{sangiao17}, to mechanically detect the magnetization dynamics in the sample placed underneath. Using piezo-electric displacement stages, this MRFM probe is set 1.6~µm above the center of the YIG disk. When SWs are excited in the sample by the microwave field, the longitudinal component of magnetization is reduced, hence the dipolar force on the MRFM probe. It results in a displacement of the cantilever beam, which is detected optically. The rf excitation applied to the sample via the antenna is pulse modulated at the mechanical resonance frequency $f_c \simeq 12.91$~kHz of the cantilever to improve the signal-to-noise ratio by its quality factor ($Q \simeq 2000$ under vacuum). A piezoelectric bimorph glued on the cantilever holder and a phase-locked loop are used to constantly drive the cantilever at resonance and fixed amplitude of 10~nm. The voltage applied on the bimorph corresponds to the MRFM signal.

\section{Micromagnetic modeling with the normal modes} \label{sec:micromag}

Micromagnetic simulations using an eigenmode solver, implemented in the micromagnetic code MaGICo \cite{magico}, are performed to calculate the SW spectrum (Fig.~\ref{fig:2}(b)) and the NLFS (Fig.~\ref{fig:4}(b)). The geometry of the body, a 52~nm thick disk of 1~µm in diameter, is discretized using a 80$\times$80$\times$1 mesh. The values of the magnetic parameters used in the simulation were those determined experimentally: saturation magnetization $M_s = 141$~kA/m, exchange constant $A_{ex} = 3.7$~pJ/m, gyromagnetic ratio $\gamma = 28$~GHz/T, damping $\alpha=5\times 10^{-4}$. The applied field in the plane is set to 27.1~mT. 

The magnetic ground state $\vec{m}_0$ is first computed for the specific geometry and applied magnetic field. The equation describing magnetization dynamics, the Landau-Lifshitz-Gilbert equation, is then linearized around the ground state. This problem can be formulated as a generalized eigenvalue problem \cite{aquino09} where $\varphi_h$ gives the precession profile of the mode with resonance frequency $\omega_h$ in the plane locally orthogonal to $\vec{m}_0$:
\begin{equation}
    \mathcal{A}_{0\perp}\varphi_h=\omega_h \mathcal{B}_0 \varphi_h \,. 
\end{equation}
The operator $\mathcal{A}_0$ is given by:
\begin{equation}
    \mathcal{A}_0\varphi_h= \mathcal{C} +h_0 \mathcal{I} \,.
\end{equation}
The operator $\mathcal{C}$ contains the energy terms of the internal effective field (exchange, dipolar, and anisotropy), and $h_0$ is given by the projection of the total effective field in the direction of equilibrium magnetization. $\mathcal{A}_{0\perp}$ is the projection of $\mathcal{A}_0$ on the plane point-wise orthogonal to $\vec{m}_0$ and $\mathcal{B}_0$ is the cross-product operator (precession). 

The eigenvectors $(\varphi_h)$ from an orthonormal basis, with the normalization convention given by:
\begin{equation}
    (\varphi_h,\mathcal{A}_{0\perp}\varphi_k)=\frac{1}{V}\int_\Omega \varphi_h^*
    \mathcal{A}_{0\perp}\varphi_k dV = \delta_{hk} \,.
\end{equation}
By projecting the dynamic magnetization on the eigenbasis, $\delta m_{\perp}=\sum_h a_h \varphi_h$, the nonlinear LLG equation can be re-written as a system of coupled equations \cite{perna22}:
\begin{multline} \label{eq:NMM}
    \dot{a}_h = j\omega_h \left( b_h+\sum_i b_{hi} a_i  + \sum_{i,j} c_{hij} a_i a_j \right. \\
    \left. + \frac{1}{2} \sum_{i,j,k} d_{hijk} a_i a_j a_k + \ldots \right) \,.
\end{multline}
The $c_{hij}$ are the 3-magnon scattering coefficients and $d_{hijk}$ the 4-magnon ones. They are given by integrals over the magnetic body computed using the mode profiles $\varphi_{h,i,j,k}$, the $\mathcal{C}$ matrix and $\vec{m}_0$. 

\begin{figure}
	\centering
        \includegraphics[width=\columnwidth]{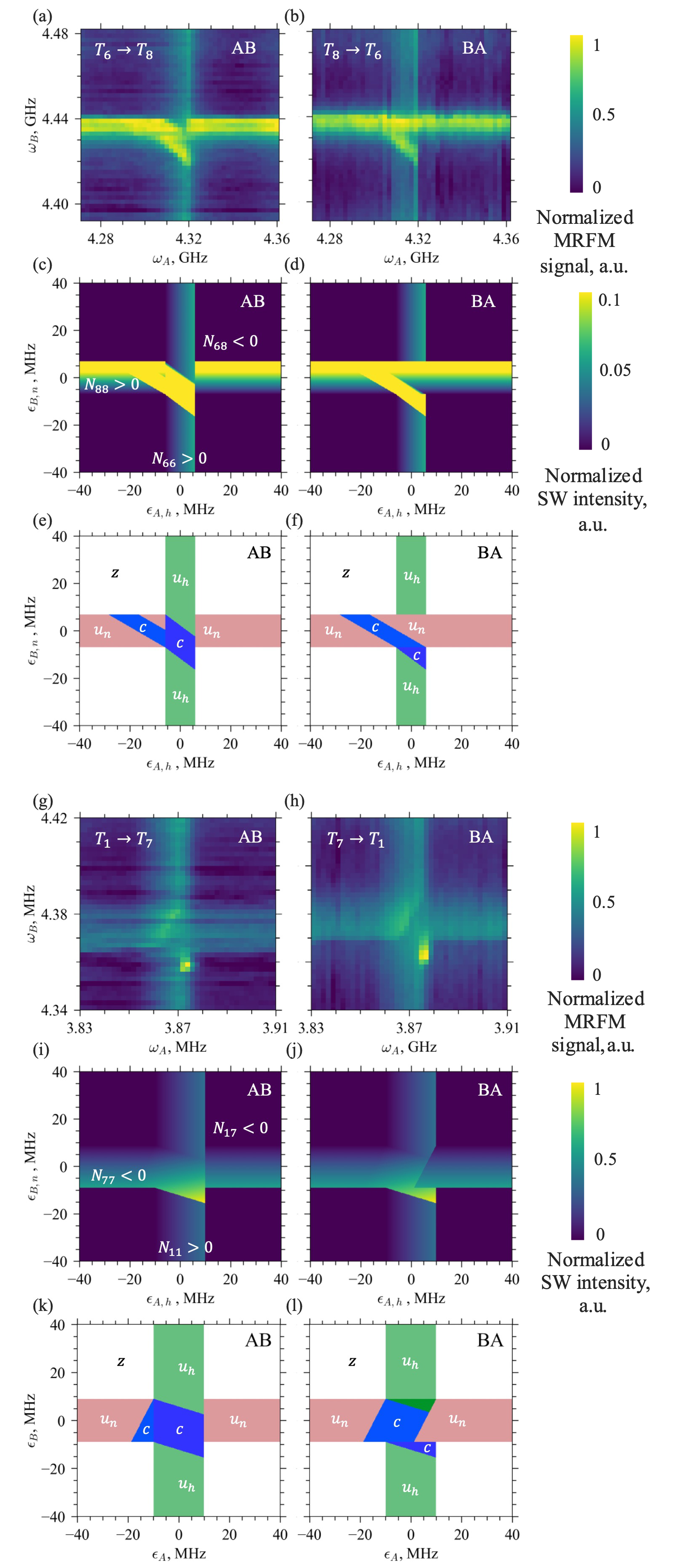}
	\caption{Additional crossings $T_6$-$T_8$ and $T_1$-$T_7$: experimental MRFM signal (a,b) and (g,h), calculated total SW intensity in the steady state (c,d) and (i,j), and associated phase diagrams (e,f) and (k,l), respectively. The labels in the phase diagrams are the same as in Fig.~\ref{fig:4}. The NFS ratios are fitted from the experiment: $N_{66}/N_{68}=-1.3 \pm 0.1~$,  $N_{88}/N_{68}=-0.6 \pm 0.1~$, $N_{11}/N_{17}=-3.5 \pm0.3~$ and $N_{77}/N_{17} = 1.8\pm0.2~$.}
	\label{fig:5}
\end{figure}

The equations for parallel pumping (\ref{eq:1t-pp}-\ref{eq:2t-pp}) are established by selecting the dominant resonant terms in (\ref{eq:NMM}). For the complete derivation, as well as the precise definition and computation of the relevant parameters, such as the NFS, the reader is referred to \cite{aquino26} that discusses in detail all aspects of the general theory.

\section{Other crossings} \label{sec:otherXings}

The analysis carried out in Fig.~\ref{fig:4} can be performed for the 36 crossings shown in Fig.~\ref{fig:3}(a). In Fig.~\ref{fig:5} we show two crossings corresponding to cases where $N_{hn} < 0$ ($T_6$-$T_8$ and $T_1$-$T_7$). As before, we fit the ratios between s- and m-NFS to reproduce the experimental crossings. Once again, for certain pumping frequencies, the steady state depends on the sequence of the microwave pulses. The phase diagrams in Fig.\ref{fig:5}(e,f,k,l), showing the steady-state solutions for our PWM experiment, allow to determine the origin of the non-commutativity. These crossings exemplify the wide variety of spectroscopic patterns observed in our system, as well as their accurate description by our theory only accounting for the s- and m-NFS.

\end{document}